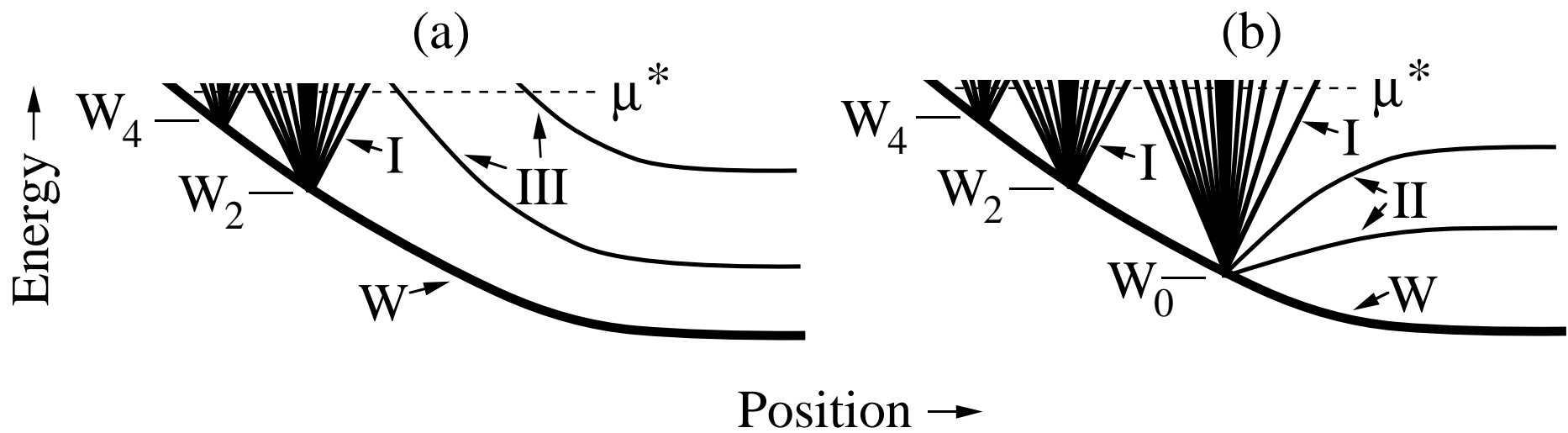

Figure 1

# Composite Fermion Theory, Edge Currents and the Fractional Quantum Hall Effect

George Kirczenow and Brad L. Johnson
Department of Physics, Simon Fraser University, Burnaby, B.C., Canada, V5A 1S6.

We present a mean field theory of composite fermion edge channel transport in the fractional and integer quantum Hall regimes. An expression relating the electro-chemical potentials of composite fermions at the edges of a sample to those of the corresponding electrons is obtained and a plausible form is assumed for the composite fermion Landau level energies near the edges. The theory yields the observed fractionally quantized Hall conductances and also explains other experimental results. We also discuss briefly some experiments that are relevant to the question whether fractional edge states in real devices should be described as Fermi or Luttinger liquids.

**Introduction**

The composite fermion theory introduced by Jain [1] and developed further by Lopez and Fradkin [2] and by Halperin, Lee and Read [3] has been remarkably successful in explaining and predicting many interesting phenomena that occur in two-dimensional electronic systems at high magnetic fields. It has provided a unified explanation [1] of the integer [4] and fractional [5] quantum Hall effects, and a systematic understanding of the relative stabilities of the fractional quantum Hall phases [1] that are observed experimentally. It also explains [1] the scaling with magnetic field of the energy gaps that are observed [6] in the fractional quantum Hall regime. Another remarkable result [3] of the composite fermion theory is that at the Landau level filling factor $\nu = 1/2$, where the effective magnetic field experienced by the composite fermions vanishes, the lowest spin-polarized electron Landau level has many features in common with systems of electrons at zero magnetic field. For $\nu \approx 1/2$ it behaves somewhat like a system of quasi-classical electrons at very low magnetic fields. These predictions have been confirmed experimentally. [7]

Since the fractional quantum Hall effect can be viewed as an integer quantum Hall effect of composite fermions, [1] an important question has been whether it is possible to construct a transport theory of the fractional quantum Hall effect based on composite fermion edge states, and thus to generalize the very successful edge state theories of the integer quantum Hall effect [8] to the fractional quantum Hall regime. Such a theory of composite fermion edge states has recently been developed at the level of mean field theory and the main ideas are outlined in this talk.

**Composite Fermion Mean Field Theory and Fractionally Quantized Hall Conductances**

In the Chern-Simons version of the composite fermion theory, a singular gauge transformation is performed which has the effect of attaching a tube of fictitious gauge flux carrying an even number of flux quanta to each electron. [1]-[3] The composite particles thus formed obey fermi statistics and are referred to as "composite fermions."[1] Many of the predictions of composite fermion theory can be understood within a mean field description in which the interactions between composite fermions that are due to the vector potentials associated with the tubes of gauge flux at-tached to the electrons, are replaced by interactions with a fictitious average magnetic field and a fictitious electric field. The fictitious magnetic field [1]-[3] is given by $\mathbf{B}_g = -mn\hat{\mathbf{B}}h/e = -m\nu\mathbf{B}$ where $m$ is the (even) number of quanta of gauge flux carried by each composite fermion, $n$ is the two-dimensional electron density, $\nu = nh/(eB)$ is the Landau level filling parameter, and $\hat{\mathbf{B}}$ is the unit vector in the direction of the true magnetic field $\mathbf{B}$. Thus the composite fermions interact with a total effective magnetic field $\mathbf{B}_{eff} = \mathbf{B} + \mathbf{B}_g$. The fictitious electric field [9] is given by $\mathbf{E}_g = -(\mathbf{J} \times \hat{\mathbf{B}})mh/e^2$, where $\mathbf{J}$ is the two-dimensional electric current density.

We assume that the electron density is slowly varying with position and that, for $\nu$ in the vicinity of $1/m$, the composite fermion Landau level energies behave qualitatively like

$$\varepsilon_{m,r} = \left(r + \frac{1}{2}\right)\hbar e |B_{eff}|/m^* + W. \tag{1}$$

Here $r = 0, 1, 2,...$ and $W$ is the position-dependent composite fermion effective potential energy. $m^*$ is the composite fermion effective mass. Equation (1) is believed to be a good description of the composite fermion Landau level structure in uniform systems [1] and yields edge states that propagate in the direction consistent with experiment [10]. The assumed behavior of the composite fermion Landau levels near an edge is shown schematically in Fig.1 for two cases of interest.

The composite fermion charge density is equal everywhere to the electron charge density because the transformation to composite fermions is a gauge transformation. Therefore the electric field felt by the composite fermions is the same as that felt by electrons except for an additional term $\mathbf{E}_g$ that does not occur in the ordinary electron picture. This term contributes to the difference between the effective electrochemical potentials $\mu^*$ of composite fermions at different edges of a current-carrying Hall bar but not to the corresponding electron electrochemical potentials $\mu$. Thus

$$\mu_i^* - \mu_j^* = \mu_i - \mu_j + \int_j^i e\mathbf{E}_g \cdot d\mathbf{r} = \mu_i - \mu_j - \int_j^i (mh/e)\mathbf{J}\cdot\hat{\mathbf{B}} \times d\mathbf{r}, \tag{2}$$

where $i$ and $j$ label different edges of the Hall bar

Since the charge of a composite fermion is $-e$, the electric current carried through a Hall bar at zero temperature by a particular composite fermion Landau level $(m,r)$ can be written

$$I_{m,r} = -\frac{e}{h}\int \frac{\partial \varepsilon_{m,r}}{\partial k} dk = -\frac{e}{h}\int d\varepsilon_{m,r} = -\frac{e}{h}\Delta\varepsilon_{m,r}, \tag{3}$$

as in edge state theories of the integer quantum Hall effect. [8] The integrals are over occupied states.

The type I edge states in Fig.1 start at the apex of a "fan" (where the composite fermion Landau levels come together at $\mathbf{B}_{eff} = 0$) and quickly pass through the Fermi level as $|\mathbf{B}_{eff}|$ increases because of the spatially varying electron density. For the type I edge states equation (3) means that $I_{m,r} = -e(\mu^* - W_m)/h$. But since $(\mu^* - W_m)$ is the local "Fermi energy," it is plausible that at fixed $\mathbf{B}$ it should depend only on the local density where $\mathbf{B}_{eff} = 0$, which is independent of $\mu^*$ since $\mathbf{B}_{eff} = 0$ where $\nu = 1/m$. Thus $\mu^* - W_m$ (and hence $I_{m,r}$) is independent of $\mu^*$. This means that the current carried by a type I mode does not change when the edge electrochemical potential changes. That is, type I modes *at edges with well-defined composite fermion electrochemical potentials* are "silent."

Now consider a sample with $\mathbf{B}_{eff}$ parallel to $\mathbf{B}$ in the bulk. The Landau levels present are of types I and III, as is illustrated in Fig.1(a). From equation (3), the current carried by a type III Landau level is $I_{m,r} = -e\Delta\mu^*/h$ where $\Delta\mu^*$ is the difference in composite fermion electrochemical po-tential across the sample. Since type I modes are silent, the net current through the sample is thus $I = -ep\Delta\mu^*/h$, where $p$ is the number of composite fermion Landau levels that are occupied in the bulk. To compare this result with experiments one must eliminate the composite fermion electrochemical potential difference $\Delta\mu^*$ in terms of the corresponding electron electrochemical potential difference $\Delta\mu$, since the latter quantity is the difference between the electrochemical potentials of the electron reservoirs attached to the sample that is actually measured. Using equation (2) which takes the form $\Delta\mu^* = \Delta\mu + mhI/e$, where $m$ is the number of quanta of gauge flux per composite fermion in the bulk, this yields $I = -e\Delta\mu p/((mp+1)h)$, or a Hall conductance $G_H = -eI/\Delta\mu = e^2p/((mp+1)h)$. Notice that *the fractional (as opposed to integer) quantized value of the Hall conductance arises from the fact that the effective electrochemical potentials of composite fermions differ from those of electrons*. The case where $\mathbf{B}_{eff}$ is antiparallel to $\mathbf{B}$ in the bulk is analyzed similarly (using the above result that $\mu^*$-$W_m$ is independent of $\mu^*$) to yield $G_H = e^2p/((mp-1)h)$.

The values $G_H = e^2p/((mp\pm1)h)$ of the Hall conductances thus obtained are just those that are found experimentally for the Landau level filling fractions $\nu = p/((mp\pm1)$ that correspond to $p$ filled composite fermion Landau levels in the bulk. That is, these results agree with experiments.

**Are Fermi or Luttinger Liquids Present at the Edges of Fractional Quantum Hall Devices?**

An important question that needs to be resolved experimentally is whether the edge states in real fractional quantum Hall devices constitute a Fermi liquid, as is suggested by the present mean field theory, or a chiral Luttinger liquid as has frequently been assumed in the literature. [11]

Experiments on selective population and detection of fractional edge channels [12] indicate that more than one conducting mode is present at the edge in $\nu = 1$ bulk systems, in contrast to the prediction of the chiral Luttinger liquid formalism that there should be only one mode present. These experiments are, however, in agreement with the present composite fermion mean field theory. [10] Also the observation of intermediate fractional plateaus (between $G = \nu e^2/h$ and $G = 0$, where $\nu$ is the Landau level filling fraction in the bulk) in the two-terminal conductance $G$ of point contacts[13] has not been explained by the Luttinger liquid theories, but has been explained by the composite fermion mean field theory.[10] However, these experiments did not probe the nature of the quasi-particles at the edge *directly*.

Measurements of the temperature dependence of the conductance of a point contact can be a direct probe, and experiments on *disordered* point contacts [14] have yielded results consistent with the predictions of chiral Luttinger liquid theory. For example, the off-resonance two-terminal conductance of a point contact was observed to decrease with decreasing temperature as a power of the temperature. However, these experiments do *not* rule out the possibility that the fractional edge states may be a Fermi liquid because the observed behavior of the conductance may have been due to the internal electronic structure of the point contact and not to exotic physics of the fractional edge states. For example, conduction via a pinned one-dimensional Wigner crystal,[15] or via one or more Coulomb-blockaded quantum dots[16] (either of which can plausibly be present in a disordered point contact at low electron densities) can also result in the conductance decreasing at low temperatures as a power of the temperature. Thus in order to establish definitively whether fractional edge states are Fermi or Luttinger liquids, further experiments are clearly needed. It would be particularly interesting to know the temperature dependence of the Aharonov-Bohm conductance resonances of a ballistic nanoscale constriction containing an antidot in the fractional regime. These resonances have already been observed. [13] Their physical origin implies that their temperature dependence should *unambiguously* reflect the nature of the quasi-particles in the fractional edge states. A composite fermion mean field theory of them and related phenomena will be presented elsewhere. [17] Preliminary measurements of their temperature dependence [18] appear to be consistent with a Fermi liquid picture of fractional edge states, but further experimental work is needed to determine how they behave at the lowest temperatures.

We wish to thank C. J. B. Ford, A. S. Sachrajda, V. J. Goldman, A. M. Chang, J. Frost, D. Loss and M. P. A. Fisher for interesting discussions.

**References:**

[1]  J. K. Jain, Phys. Rev. Lett. **63**, 199 (1989); Science **266**, 1199 (1994).
[2]  A. Lopez and E. Fradkin, Phys. Rev. B**44**, 5246 (1991).
[3]  B. I. Halperin, P. A. Lee and N. Read, Phys. Rev. B**47**, 7312 (1993).
[4]  K. von Klitzing, G. Dorda and M. Pepper, Phys. Rev. Lett. **45**,494 (1980).
[5]  D. C. Tsui, H. L. Störmer and A. C. Gossard, Phys. Rev. Lett. **48**,1559 (1982).
[6]  R. R. Du *et al*., Phys. Rev. Lett **70**, 2944 (1993); I. V. Kukushkin *et al*., Phys. Rev. Lett **72**, 736 (1994); H. C. Manoharan *et al*., Phys. Rev. Lett **73**, 3270 (1994);D. R. Leadley *et al*., Phys. Rev. Lett **72**, 1906 (1994).
[7]  R. L. Willet *et al*., Phys. Rev. Lett **71**, 3846 (1993); W. Kang *et al*., Phys. Rev. Lett **71**, 3850 (1993); V. J. Goldman *et al*., Phys. Rev. Lett **72**, 2065 (1994).
[8]  B. I. Halperin, Phys. Rev. B**25**, 2185 (1982); P. Streda, J. Kucera and A. H. MacDonald, Phys. Rev. Lett. **59**,1973(1987); J. K. Jain and S. A. Kivelson, Phys. Rev. Lett. **60**,1542 (1988); M. Büttiker, Phys. Rev. B**38**, 9375 (1988).
[9]  S. C. Zhang, H. Hansson and S. A. Kivelson, Phys. Rev. Lett. 62, 82 (1989); S. C. Zhang, Int. J. Mod. Phys. B**6**, 25 (1992); B. Rejaei and C. W. J. Beenakker, Phys. Rev. B**43**, 11392 (1991), Phys. Rev. B**46**, 15566 (1992); A. S. Goldhaber and J. K. Jain, Phys. Lett. A**199**, 267 (1995).
[10]  G. Kirczenow and B. L. Johnson, Phys. Rev. B**51**,17579(1995). See paper cond-mat/ 950258, www site http://xxx.lanl.gov/ for legible figures.
[11]  See C. L. Kane and M. P. A. Fisher, Phys. Rev. B**51**, 3449 (1995) and references therein.
[12]  L. P. Kouwenhoven, *et al*., Phys.Rev. Lett. **64**, 685 (1990); A. M. Chang and J. E. Cunningham, Phys. Rev. Lett **69**, 2114 (1992); A. S. Sachrajda private communication.
[13]  C. J. B. Ford, *et al*., J. of Phys. Condensed Matter **6**, L725 (1994); V. J. Goldman and B. Su, Science **267**, 1010 (1995); J. D. F. Franklin *et al*. Proc. EP2DSXI, to be published in Surface Science.
[14]  F. P. Milliken, C. P. Umbach and R. A. Webb, Physics Today Vol. **47**, No 6, p. 21 (1994).
[15]  L.I.Glazman, I. M. Ruzin and B. I. Shklovskii, Phys. Rev. B**45**, 8454 (1992).
[16]  K.A.Matveev, L.I.Glazman and H.U.Baranger; A. Furusaki and K. A. Matveev, preprints.
[17]  G. Kirczenow, unpublished.
[18]  C. J. B. Ford, private communication; J. D. F. Franklin *et al*. Proc. EP2DSXI, to be published in Surface Science; V. J. Goldman, private communication.

**Figure Captions**

**Fig.1** Schematic drawing of composite fermion Landau level structure near the edge for $\mathbf{B}_{\text{eff}}$ in the bulk parallel (case (a)) and antiparallel (case (b)) to $\mathbf{B}$. The apex of each "fan" of energy levels occurs where $\nu = 1/m$, and there $\varepsilon_{m,r} = W = W_m$. The three different types of composite fermion edge modes are labelled I, II and III. $R_H = 2e^2/h$ and $2e^2/(3h)$ in cases (a) and (b) respectively.